\documentclass[pra, twocolumn,a4paper,nofootinbib,superscriptaddress]{revtex4}
\pdfoutput=1
\usepackage{hyperref}
\usepackage{graphicx}
\usepackage{amsmath}
\usepackage{amssymb}
\usepackage{amsthm,dsfont}
\usepackage{color}
\usepackage{natbib}
\hypersetup{colorlinks=true, urlcolor=blue, linktocpage}

\theoremstyle{plain}

\newcommand{\abs}[1]{\left|#1\right|}

\newcommand{\bra}{\langle}
\newcommand{\ket}{\rangle}
\newcommand{\bq}{\begin{equation}}
\newcommand{\eq}{\end{equation}}
\newcommand{\ba}{\begin{align}}

\definecolor{red}{rgb}{0.9,0,0}
\definecolor{green}{rgb}{0,0.8,0}
\definecolor{blue}{rgb}{0,0,0.8}
\definecolor{cautionred}{rgb}{1.0,0,0}

\definecolor{maroon}{rgb}{0.7,0,0}

\definecolor{ngreen}{rgb}{0.3,0.7,0.3}

\definecolor{golden}{rgb}{0.8,0.6,0.1}

\begin{document}
\bibliographystyle{apsrev4-1}

\title{All two-qubit states that are steerable via Clauser-Horne-Shimony-Holt-type
	correlations are Bell nonlocal}
\author{Parth Girdhar}
\affiliation{Centre for Engineered Quantum Systems, School of Physics, The University of Sydney, Sydney, NSW 2006, Australia}
\author{Eric G. Cavalcanti}
\affiliation{Centre for Quantum Dynamics, Griffith University, Brisbane, QLD 4111, Australia}
\affiliation{School of Physics, The University of Sydney, Sydney, NSW 2006, Australia}

\begin{abstract}
We derive an inequality that is necessary and sufficient to show Einstein-Podolsky-Rosen (EPR) steering in a scenario employing only correlations between two arbitrary dichotomic measurements on each party. Thus the inequality is a complete steering analogy of the Clauser-Horne-Shimony-Holt (CHSH) inequality, a generalisation of the result of Cavalcanti {\it et al}. [E. G. Cavalcanti, C. J. Foster, M. Fuwa, and H. M. Wiseman, \href{https://dx.doi.org/10.1364/JOSAB.32.000A74}{JOSA B, 32, A74 (2015)}]. We show that violation of the inequality only requires measuring over equivalence classes of mutually unbiased measurements on the trusted party and that in fact assuming a general two qubit system arbitrary pairs of distinct projective measurements at the trusted party are equally useful. Via this it is found that for a given state the maximum violation of our EPR-steering inequality is equal to that for the CHSH inequality, so all states that are EPR steerable with CHSH-type correlations are also Bell nonlocal.
\end{abstract}

\maketitle
Einstein-Podolsky-Rosen-steering (EPR) characterizes the apparent ability to nonlocally affect a quantum state, the central problem in the infamous EPR argument~\cite{einstein1935can}, that aimed to show that quantum mechanics is incomplete. That argument considered an entangled state shared between two distant parties, and proceeded to show that by measuring one or another of two non-commuting observables on the local system, the distant system is left in different possible sets of quantum states, an effect that Schr\"{o}dinger later termed ``steering''~\cite{schrodinger1935discussion}. These allow the experimenter to predict the result of measuring one or another of two non-commuting observables at the distant system. But since the systems no longer interact, EPR argued, the local choice of measurement cannot affect the ``elements of reality'' associated with the distant system. Thus both quantities should have simultaneous reality, which EPR believed would be described by a theory more complete than quantum mechanics. However, the possibility of such a local hidden-variable (LHV) description was ruled out by Bell in 1964~\cite{bell1964einstein}.

In 1989 Reid derived variance-inequalities that are violated with EPR correlations for continuous variable systems \cite{reid1989demonstration} and this was extended to discrete variables in \cite{cavalcanti2009spin}. Wiseman, Jones and Doherty (WJD) introduced a notion of steering as the inability to construct a local hidden state (LHS) model to explain the probabilities of measurement outcomes~\cite{wiseman2007steering}. In quantum-information terms, EPR steering can be defined as the task for a referee to determine whether two parties share entanglement, when one of the parties is untrusted and using only classical communication~\cite{wiseman2007steering}. Based on this, EPR-steering inequalities were defined in \cite{cavalcanti2009experimental}, with the property that violation of any such inequality implies steering. It was shown in \cite{wiseman2007steering} and \cite{jones2007entanglement} that the set of steerable states, that is states for which there exist local measurements that produce violation of a steering inequality, are strictly a subset of entangled states and a superset of states that violate a Bell inequality (Bell-nonlocal states). In particular the set of Bell-local Werner states that are unsteerable was found but no clear connection between the set of mixed steerable states and Bell nonlocal states has been determined.  Experiments on entangled photon pairs \cite{ou1992realization, bowen2003experimental, howell2004realization, saunders2010experimental, saunders2012simplest, smith2012conclusive, bennet2012arbitrarily, handchen2012observation, schneeloch2013violation, steinlechner2013strong, fuwa2015experimental} have produced violations of steering inequalities thus demonstrating the EPR paradox; in particular \cite{wittmann2012loophole} reported loophole-free steering inequality violation, analogous to the much sought-after loophole-free Bell inequality violation, that was reported for the first time only this year \cite{hensen2015loophole}. The WJD formalism has also had application in quantum information theoretic tasks such as one-sided device-independent quantum key distribution, quantum teleportation and subchannel discrimination \cite{branciard2012one, reid2013signifying, piani2015necessary}.

Recently, the authors of \cite{cavalcanti2015analog} derived an EPR-steering analogue of the Clauser-Horne-Shimony-Holt (CHSH) inequality, that is, an inequality that is necessary and sufficient to demonstrate EPR steering in a scenario involving only correlations between two dichotomic measurements on each subsystem. However, this inequality requires that measurements by the trusted party (the ``steered'' party) be mutually unbiased. Here we produce a necessary and sufficient steering inequality in the same CHSH scenario as \cite{cavalcanti2015analog}, that applies to any pair of projective measurements at the trusted party. This is presented in Sec. II with a full proof in Appendix A. In Appendix B the set of unsteerable correlations for arbitrary dichotomic positive operator-valued measures (POVM's) is found though a simple necessary and sufficient inequality cannot be constructed for this case. In Sec. III it is shown that the inequality is violated if and only if an inequality involving mutually unbiased measurements is also violated. This fact is used in Sec. IV to find the maximum violation of this EPR-steering inequality for a given bipartite state, which turns out to be equal to the maximum violation of the CHSH inequality for the given state as calculated in \cite{horodecki1995violating}. The inequalities have the same right hand side hence we find an equivalence between steering and nonlocality for this scenario. Thus the known distinction between the sets of Bell-nonlocal and steerable states cannot be determined with CHSH-type correlations alone. 

\section{Necessary and Sufficient Steering Inequality} - Here we develop the EPR-steering formalism, following the notation of \cite{cavalcanti2015analog}, and develop the necessary and sufficient EPR-steering inequality for the CHSH scenario with a full proof in Appendix A. Through a similar process the boundary of the set of unsteerable correlations can be found for dichotomic POVM's as we show in Appendix B. We have a pair of isolated systems, one at Alice and the other at Bob. We denote a measurement at Alice's (Bob's) system as $A$ ($B$), chosen from a the set of observables $\mathfrak{D_\alpha}$, ($\mathfrak{D_\beta}$) in the Hilbert space of Alice's (Bob's) system, with outcomes labelled by $a\in\mathfrak{L_\alpha}(b\in\mathfrak{L_\beta})$. 
A state \emph{W} shared between Alice and Bob is defined as \emph{Bell local} or it has a LHV model if and only if it is the case that $\forall a, b, A,B$ the joint probability distributions can be written in the form: 
\begin{equation}
	P(a, b|A,B;W)=\sum_\lambda\wp(\lambda)\wp(a|A,\lambda)\wp(b|B,\lambda)
\end{equation}
where $\wp(\lambda)$ is a probability distribution over hidden variables $\lambda \in \Lambda$, $\wp(a|A,\lambda)$ is the probability of outcome $a$ for measurement $A$ given $\lambda$, and likewise for $\wp(b|B,\lambda)$.

A state \emph{W} is \emph{unsteerable} or it has a \emph{local hidden variable -- local hidden state} (LHV-LHS) model if and only if all joint distributions have the form: 
\begin{equation} \label{eq:LHV-LHS}
	P(a, b|A,B;W)=\sum_\lambda\wp(\lambda)\wp(a|A,\lambda)P(b|B;\rho_\lambda)
\end{equation}
where now it is further assumed that $\lambda$ determines a local quantum state $\rho_\lambda$ for Bob, and $P(b|B;\rho_\lambda)$ is the quantum probability of outcome $b$ if $B$ is measured on $\rho_\lambda$. Since those probabilities are given by a quantum state, they must be constrained by uncertainty relations. 

This scenario has an operational meaning: Bob wishes to verify if $W$ is entangled given joint distributions of outcomes between Alice's and Bob's measurements, but assuming only Bob's outcomes are ``trusted'' as arising from quantum measurements. Here it is not possible to determine entanglement via state tomography as only Bob's measurements are trusted, however showing that not all distributions can be expressed as Eq. \eqref{eq:LHV-LHS} is sufficient to show entanglement. The scenario in which both Alice's and Bob's measurements are untrusted would require testing the joint distributions for Bell nonlocality. 

The set of correlations in Eq. \eqref{eq:LHV-LHS} forms a convex set \cite{cavalcanti2009experimental} so we can express it in terms of its extreme points as: 
\begin{equation}
P(a, b|A,B)=\sum_{\chi}\int d\xi\wp(\chi,\xi)\delta_{a, f(A,\chi)}\langle\psi_\xi|\Pi_{b}^{B}|\psi_\xi\rangle
\end{equation}
where $\Pi_{b}^{B}$ is a projector for outcome $b$ of measurement $B$, $\chi$ is a parameter that determines all values of $A$ via a function $f(A,\chi)$ and $\xi$ determines a pure state $\psi_\xi$ for Bob.

In constructing an EPR-steering inequality analogous to the CHSH inequality, we assume Alice and Bob can choose between two measurements $\{A,A'\}$ and $\{B,B'\}$ respectively, with possible outcomes $a,b \in \{1,-1\}$. We consider the ordered set of correlations $(\langle{AB}\rangle, \langle{A'B}\rangle, \langle{AB'}\rangle, \langle{A'B'}\rangle)$ obtained in such an experiment, where $\langle{AB}\rangle=P(a=b|A,B)-P(a=-b|A,B)$ and similarly for the other terms. These are the same correlations appearing in the CHSH inequality, and we want to ask what we can say about the steerability of a state using only this information. A LHV-LHS model can reproduce these correlations if and only if there exists a probability distribution $\wp(\chi,\xi)$ such that they can be expressed as:
\begin{equation}\label{eq:LHV-LHScorr}
\langle{AB}\rangle=\sum_{\chi}\int d\xi\wp(\chi,\xi)(2p_1^{A}(\chi)-1)(2p_1^{B}(\xi)-1) \;,
\end{equation}
where $p_1^A(\chi)=\wp(1|A,\chi)$ and $p_1^B(\xi)=P(1|B,\psi_\xi)$.  

For Alice there are four extreme values of $\chi$, which we label as $\chi\in\{1,2,3,4\}$ corresponding respectively to $p_1^A=p_1^{A'}=1$, $p_1^A=1-p_1^{A'}=1$, $p_1^A=1-p_1^{A'}=0$, $p_1^A=p_1^{A'}=0$. $\{B,B'\}$ are quantum projective measurements, which can be written as $B=2\, \Pi_1^B - I$, where $\Pi^B_1$ is the projector onto the $+1$ eigenstate of $B$, and similarly for $B'$. Following \cite{cavalcanti2015analog}, let $\mu=$Tr$\{\Pi_1^{B}\Pi_1^{B'}\}$ and the possible pairs of probabilities $(p_1^B(\xi), p_1^{B'}(\xi))$ with $p_1^B(\xi)=\langle\psi_\xi|\Pi_{1}^{B}|\psi_\xi\rangle$ form an ellipse, which can be parameterised as:
\begin{eqnarray}\label{eq:ellipse1}
2p_{1}^{B}(\xi)-1 & = &\cos(\xi+\beta)\\
2p_{1}^{B'}(\xi)-1 & = &\cos(\xi-\beta).
\end{eqnarray}
where $\beta = \arctan (\frac{ \sqrt{1-\mu}}{\sqrt{\mu}} )$ with $0\leq\beta\leq\pi/2$. It turns out that if $\{B,B'\}$ are dichotomic POVM's then $(p_1^B(\xi), p_1^{B'}(\xi))$ also form an ellipse, as we show in Appendix B, and so via a proof similar to that presented here the set of unsteerable correlations can be found for POVM's but an analogous inequality does not exist. Varying $\xi$ and $\chi$, the possible  values for the integrands in Eq. \eqref{eq:LHV-LHScorr} for each correlation, are given by:
\begin{equation}
\begin{array}{ccccc}
 & \chi=1 & \chi=2\\
\left\langle AB\right\rangle &\cos(\xi+\beta) &\cos(\xi+\beta) \\
\left\langle A'B\right\rangle &\cos(\xi+\beta) & -\cos(\xi+\beta) \\
\left\langle AB'\right\rangle &\cos(\xi-\beta) &\cos(\xi-\beta) \\
\left\langle A'B'\right\rangle&\cos(\xi-\beta) & -\cos(\xi-\beta)
\end{array}.\label{eq:extreme points}
\end{equation}

The correlations for $\chi=3$($4$) can be obtained from those for $\chi=1$($2$) by making $\xi \rightarrow \xi+ \pi$, and so it's sufficient to consider $\chi=1,2$. Thus the vector of correlations has a LHV-LHS model if and only if they can be written as a convex combination of the vectors given by the columns on Eq.~\eqref{eq:extreme points}. Let $C_1$ be the convex hull of the $\chi=1$ column of Eq.~\eqref{eq:extreme points} and $C_2$ that for $\chi=2$.

Then the set $C$ of all vectors of correlations in which each correlation is of the form Eq. \eqref{eq:LHV-LHScorr} is the convex hull of the union of $C_1$ and $C_2$. In the basis:
\begin{align}\label{eq:basis}
\mathbf{e}_{1} & =(1,1,0,0)\nonumber \\
\mathbf{e}_{2} & =(0,0,1,1)\nonumber \\
\mathbf{e}_{3} & =(1,-1,0,0)\nonumber \\
\mathbf{e}_{4} & =(0,0,1,-1),
\end{align}
the vectors making up the boundaries of $C_1$ and $C_2$ have form $\cos(\xi+\beta)\mathbf{e}_{1}+\cos(\xi-\beta)\mathbf{e}_{2}$ and $\cos(\xi+\beta)\mathbf{e}_{3}+\cos(\xi-\beta)\mathbf{e}_{4}$ respectively.

Now the curve (\emph{x,y})=(cos($\xi+\beta$),cos($\xi-\beta$)) is an ellipse which can also be expressed as
\begin{align}\label{eq:quadraticproj}
x^2+y^2-2xycos(2\beta)=sin^2(2\beta) 
\end{align}

This leads to the conjecture that for $\mathbf{v}=(v_{1},v_{2},v_{3},v_{4})$ in the basis $\{\mathbf{e}_{i}\}$, we have $\mathbf{v}\in C$ if and only if 
\begin{multline}\label{eq:main_vbasis}
\frac{1}{sin(2\beta)}(\sqrt{v_1^2+v_2^2-2v_1v_2cos(2\beta)}+\\\sqrt{v_3^2+v_4^2-2v_1v_2cos(2\beta)})\leq1 
\end{multline}

In the original basis this is 
\begin{equation}\label{eq:main}
\frac{1}{sin(2\beta)}(\sqrt{u_1}+\sqrt{u_2})\leq2
\end{equation}
where 
\begin{align}\label{eq:u1u2}
u_1 =& \left\langle (A+A')B\right\rangle^{2} + \left\langle (A+A')B'\right\rangle^{2}\nonumber\\
  &- 2 \cos(2\beta) \left\langle (A+A')B\right\rangle \left\langle (A+A')B'\right\rangle\\
u_2 =&\left\langle (A-A')B\right\rangle ^{2}+\left\langle (A-A')B'\right\rangle ^{2}\nonumber\\
  &-2 \cos(2\beta) \left\langle (A-A')B\right\rangle\left\langle (A-A')B'\right\rangle
\end{align}

In other words, Eq. \eqref{eq:main} is the necessary and sufficient inequality for the four correlations considered to have a LHV-LHS models, for arbitrary measurements on Bob's side. It is thus an analog of the CHSH inequality for EPR steering. It reduces to Eq. (21) in \cite{cavalcanti2015analog} for $\beta=\frac{\pi}{4}$ which corresponds to $\mu=0.5$. The full proof of this conjecture is in Appendix A.

\section{Equivalence Classes of Measurements}- Whilst arbitrary dichotomic projective measurements can be made on Bob's side, there are actually equivalence classes of measurements $B'$, for fixed $A, A'$, and $B$, for which all $B'$ in the same class result in the same left hand side of Eq.~\eqref{eq:main}, as we will now show. Each equivalence class can be associated with a measurement mutually unbiased to $B$. Then optimising the inequality over measurements, for a given state, only requires optimising over mutually unbiased measurements by Bob. 
Bob's measurements are trusted and thus in accordance with quantum mechanics they are Hermitian operators, and by convention have $\pm 1$ eigenvalues. $B'$ can then be expressed as 
\begin{align}
B'=(2\mu-1)B+2\sqrt{\mu}\sqrt{1-\mu}B'' 
\end{align}
where \emph{B''} is an operator mutually unbiased to \emph{B}. Since $\cos(2\beta)=2\mu-1$ and $\sin(2\beta)=2\sqrt{\mu}\sqrt{1-\mu}$,
we can rewrite $u_1$ and $u_2$ in Eq.~\eqref{eq:main} in terms of \emph{A, A', B, B''} as:

\begin{align}
u_1 = &\sin^2(2\beta) \left( \langle(A+A')B\rangle^{2} + \langle(A+A')B''\rangle^{2} \right) \\
u_2 = & \sin^2(2\beta) \left( \langle(A-A')B\rangle^{2} + \langle(A-A')B''\rangle^{2} \right) \;.
\end{align}
Substituting in Eq.~\eqref{eq:main} we obtain:
\begin{multline}\label{eq:unbiased}
\sqrt{\left\langle (A+A')B\right\rangle^{2}+\left\langle (A+A')B''\right\rangle^{2}}\\
+\sqrt{\left\langle (A-A')B\right\rangle^{2}+\left\langle (A-A')B''\right\rangle^{2}}
\leq2
\end{multline}

This is equivalent to Eq.~\eqref{eq:main} for measurements $A, A', B, B''$ with $\beta=\frac{\pi}{4}$, as should be for mutually unbiased measurements $B, B''$.

Hence, if an arbitrary set of dichotomic variables $\{A, A'\}$ by Alice and $\{B, B'\}$ by Bob is measured and Bob's measurements are trusted, the four correlations between variables $\{A, A'\}$ and $\{B, B'\}$ are consistent with a LHV-LHS model if and only if the four correlations between variables $\{A, A'\}$ and $\{B, B''\}$ are consistent with a LHV-LHS model, where \emph{B''} is the mutually unbiased measurement to \emph{B} determined by B''=$\frac{B'-(2\mu-1)B}{2\sqrt{\mu}\sqrt{1-\mu}}$.

So the demonstration of steering in this scenario implies violation of Eq. \eqref{eq:main} for some pair of mutually unbiased measurements by Bob.  Equation ~\eqref{eq:main} implicitly contains $\mu$ as a variable, which depends on \emph{B} and \emph{B'} set by the experimentalist, but the equivalent inequality \eqref{eq:unbiased} does not depend on $\mu$. Given a \emph{B}, each \emph{B'} is mapped to a particular \emph{B''} (mutually unbiased to \emph{B}) and the independence of $\mu$ means that each \emph{B''} defines an equivalence class containing an infinity of \emph{B'} observables each mapped to \emph{B''}. Then the inequality Eq.~\eqref{eq:main} for a particular \emph{B'} is not only equivalent to Eq. \eqref{eq:unbiased} but is equivalent to an infinity of inequalities involving the same \emph{A, A',B} and some \emph{B'} from the equivalence class to which \emph{B'} belongs.

\section{States Steerable via CHSH-type measurements are Nonlocal}- We now show that if a two-qubit quantum state violates the steering inequality \eqref{eq:unbiased} for some set of measurements then it also violates the CHSH inequality, possibly with another set of measurements. It was shown above that Eq. \eqref{eq:unbiased} is violated by some quantum state if and only if the steering inequality \eqref{eq:main}, for general measurements of the type in the CHSH scenario, is also violated. This means, since the inequality is necessary and sufficient, that a state demonstrates steering via general CHSH-type correlations if and only if it violates the CHSH inequality. Therefore all states that demonstrate steering via CHSH-type correlations are Bell nonlocal. 

Every bipartite state  involving two qubits can be written in the form:
\begin{equation}\label{eq:bipartite} 
\rho = \frac{1}{4}\Bigg(I \otimes I+\mathbf{r\cdot \sigma}\otimes I+I\otimes \mathbf{s\cdot\sigma}+\sum_{n,m=1}^{3}t_{m n}\sigma_n\otimes\sigma_m\Bigg)
\end{equation}
where, in the notation of \cite{horodecki1995violating}, $I$ is the identity operator, $\{\sigma_n\}_{n=1}^{3}$ are Pauli matrices, $\boldsymbol{\mathbf{r}}$ and $\boldsymbol{\mathbf{s}}$  are vectors in $\mathbb{R}^3$, and $\mathbf{r\cdot\sigma}=\sum_{i=1}^{3}r_i\sigma_i$, $t_{mn}=\mathrm{Tr}(\rho\sigma_n\otimes \sigma_m)$ forms a matrix denoted $T_\rho$.

We seek to find the maximum value of the steering inequality \eqref{eq:unbiased} for this state. Unlike the CHSH inequality this steering inequality is nonlinear so its left hand side can not be replaced by the expectation value of a single operator. Defining $A=\widehat{\mathbf{a}}\cdot\sigma$, $A'=\widehat{\mathbf{a}}'\cdot\sigma$, $B=\widehat{\mathbf{b}}\cdot\sigma$, $B'=\widehat{\mathbf{b}}'\cdot\sigma$, where $\mathbf{\widehat{a}}$, $\mathbf{\widehat{a}}'$, $\mathbf{\widehat{b}}$, $\mathbf{\widehat{b}}'$ are unit vectors in $\mathbb{R}^3$, the left hand side of Eq. \eqref{eq:unbiased} can be written in the form:
\begin{multline}\label{eq:E_steer}
{E}_{Steer}=\sqrt{(\widehat{\mathbf{b}}, T_\rho(\mathbf{\widehat{a}+\widehat{a}'}))^2 +(\mathbf{\widehat{b}'}, T_\rho(\mathbf{\widehat{a}+\widehat{a}'}))^2}\\
+\sqrt{(\widehat{\mathbf{b}}, T_\rho(\mathbf{\widehat{a}-\widehat{a}'}))^2 +(\mathbf{\widehat{b}'}, T_\rho(\mathbf{\widehat{a}-\widehat{a}'}))^2}
\end{multline}

Defining orthonormal vectors $\mathbf{\widehat{c}}$, $\mathbf{\widehat{c'}}$ by:
\begin{align}
\mathbf{\widehat{a}}+\mathbf{\widehat{a}}'=2\cos\theta\mathbf{\widehat{c}}\notag\\
\mathbf{\widehat{a}}-\mathbf{\widehat{a}}'=2\sin\theta\mathbf{\widehat{c}'}   
\end{align}
where $\theta\in [0,\frac{\pi}{2}]$ we can express Eq. \eqref{eq:E_steer} as:
\begin{equation}\label{eq:E_steer2}
{E}_{Steer}=2(\cos\theta\sqrt{\left \| T_\rho\mathbf{\widehat{c}} \right \|^2}+\sin\theta\sqrt{\left \| T_\rho\mathbf{\widehat{c}}' \right \|^2})\;,
\end{equation}
where Pythagoras' Theorem has been used on the orthogonal components of $T_\rho(\mathbf{\widehat{c}})$ and $T_\rho(\mathbf{\widehat{c'}})$ in the $\mathbf{\widehat{b}}$ and $\mathbf{\widehat{b'}}$ directions. The disappearance of $\mathbf{\widehat{b}},\mathbf{\widehat{b'}}$ in the expression \eqref{eq:E_steer2} shows that the left hand side of Eq. \eqref{eq:unbiased} is independent of measurements on Bob's side, assuming the inequality applies to two qubits. Ultimately this means that verifying steering using two fixed measurements on Alice's side only requires choosing any pair of different measurements on Bob's side. 

Maximising ${E}_{Steer}$ we find:
\begin{align}\label{eq:maxE_steer}
\max({{E}_{Steer}})=&\max_{\mathbf{\widehat{c}},\mathbf{\widehat{c}'}, \theta} \left\{ 2(\cos\theta\sqrt{\left \| T_\rho\mathbf{\widehat{c}} \right \|^2}+\sin\theta\sqrt{\left \| T_\rho\mathbf{\widehat{c}'} \right \|^2}) \right\} \nonumber\\
=&\max_{\mathbf{\widehat{c}},\mathbf{\widehat{c}'}} \left\{ 2\sqrt{\left \| T_\rho\mathbf{\widehat{c}} \right \|^2+\left \| T_\rho\mathbf{\widehat{c}'} \right \|^2} \right\} \;.
\end{align}

In the last step above we maximise over angle $\theta$ keeping fixed $\mathbf{\widehat{c}},\mathbf{\widehat{c'}}$, and the optimal angle is
\begin{equation}
\theta_{\mathrm{max}}=\tan^{-1}\frac{\left \| T_\rho\mathbf{\widehat{c}'}_{\mathrm{max}} \right \|}{\left \| T_\rho\mathbf{\widehat{c}}_{\mathrm{max}} \right \|}   
\end{equation}
where $\mathbf{\widehat{c}}_{\mathrm{max}}$ and $\mathbf{\widehat{c}'}_{\mathrm{max}}$ are the vectors that maximise $\left \| T_\rho\mathbf{\widehat{c}} \right \|^2+\left \| T_\rho\mathbf{\widehat{c}'} \right \|^2$.  This is exactly the maximum of the CHSH inequality calculated in \cite{horodecki1995violating}. Since both inequalities have a right hand side of $2$, a state $\rho$ violates our CHSH-type steering inequality if and only if it also violates the CHSH inequality, possibly for different sets of measurements. Explicitly, this violation occurs iff the sum of the squares of the largest eigenvalues of $T_{\rho}$ is greater than 1 \cite{horodecki1995violating}. As an example, a Werner state $W^{\eta}$ violates the CHSH inequality for $\eta>\eta_{CHSH}=\frac{1}{\sqrt{2}}$~\cite{horodecki1995violating}, and therefore this result implies that it is steerable under CHSH correlations above this same threshold for $\eta$, confirming the result shown in \cite{roy2015optimal} (however there it was not demonstrated that on Bob's side only mutually unbiased bases need to be considered). This ``equivalence'' between steering and Bell nonlocality applies generally to dichotomic POVM measurements by Bob since the optimal measurements to show steering are some projective measurements (as a dichotomic POVM can be regarded as being a classically post-processed projective measurement \cite{d2005classical}).

\section{Discussion} The connection between steering and Bell nonlocality shown above is surprising since it was established in the seminal papers on the subject \cite{wiseman2007steering}, \cite{jones2007entanglement} that steerable states are a strict subset of Bell nonlocal states. While all pure entangled states are Bell nonlocal, and hence also steerable, a strict hierarchy exists between entangled, steerable and Bell nonlocal mixed states in general. But we see that if CHSH-type correlations demonstrate that a state is steerable then that state must also be Bell nonlocal so in a sense the hierarchy is collapsed for these types of correlations. Furthermore the independence of the left-hand side of the steering inequality on Bob's measurements means it is only necessary to vary over Alice's measurements to verify that the state is Bell nonlocal using this technique. 

We also note that if we can demonstrate steering in one direction, then the state is Bell nonlocal, and therefore steering can also be demonstrated in the other direction. Thus there is no one-way steering in this scenario, in the sense of \cite{bowles2014one}. It was shown in \cite{saunders2010experimental} both theoretically and via an optical experiment that there exist Bell-local Werner states which violate a steering inequality involving three dichotomic measurements on either side. More recently, Bowles {\it et al.} \cite{Bowles2016} have shown that some Bell-local states are one-way steerable with two projective measurements at the untrusted site and tomographic measurements on the trusted site. 

The present results would suggest that indeed at least three measurements at the trusted site are required for one-way steering. However, our inequality is necessary and sufficient when using the correlation data only---it is known that for some two-qubit states which are not detected by this inequality, steering can be detected using also the information about the marginals, not available from only the correlations~\cite{Cyril}. The question then becomes whether some of those steerable states are also Bell-local and one-way steerable. It would be interesting to derive an inequality that is necessary and sufficient for this general case where the marginals are also taken into account. Further work could explore necessary and sufficient steering inequalities that involve more than two measurement variables on each party or more than two outcomes for each measurement. 

The results above provide a partial answer to this fundamental question: for a given measurement scenario (i.e. number of parties, settings and outcomes) what are the optimal measurements to verify if an arbitrary quantum state is steerable? For the CHSH-type scenario the answer to the question above is: any pair of distinct arbitrarily chosen measurements on Bob's side and the measurements made by Alice corresponding to $\mathbf{\widehat{a}}_{max}$ and $\mathbf{\widehat{a'}}_{max}$ constructed from $\theta_{max}$ and the $\mathbf{\widehat{c}}_{max}$ and $\mathbf{\widehat{c'}}_{max}$ that maximise $\left \| T_\rho\mathbf{\widehat{c}} \right \|^2+\left \| T_\rho\mathbf{\widehat{c'}} \right \|^2$ (eigenvectors of $T_{\rho}T$ \cite{horodecki1995violating}). Recently in \cite{roy2015optimal} a computational optimisation over measurements on both sides was performed to find the maximum violation of Eq. \eqref{eq:unbiased} over the space of bipartite pure entangled qubits and amount of violation of the inequality over a class of Werner states, but we now see that it would have been sufficient to keep measurements on Bob's side fixed. This independence on Bob's measurements removes a major challenge in achieving practical nonlocality witnesses; for example in \cite{smith2012conclusive} the demonstration of steering inequality violation required substantial steps to account for non-mutually unbiased measurements. 

For pure states the connection between joint measurability (compatibility) of Alice's observables and Bell-nonlocality was examined in \cite{wolf2009measurements} and recently extended to steering in \cite{quintino2014joint, uola2014joint}. The latter works suggest that two measurements at Alice are incompatible if and only if they can be used to demonstrate steering, while the former work suggests that two incompatible measurements enable 
CHSH inequality violation. The findings in our paper also show a kind of independence from measurements on Bob's side (they are only required to be incompatible) and it is interesting that our approach via steering inequalities gives similar insights to their derivations using a reduced-state/``assemblage'' picture. But as we allow the state to be mixed, so we include entangled unsteerable states, the criterion on Alice's measurements to demonstrate steering with CHSH correlations is stronger than incompatibility, specifically that Eq. \eqref{eq:maxE_steer} must be larger than $2$. 

In conclusion we have produced a general necessary and sufficient steering inequality for CHSH-type correlations on two qubits. The violation of the inequality for a given set of measurements implies the violation of an inequality with mutually unbiased measurements on the ÔtrustedÕ side hence only mutually unbiased measurements need to be examined, as in [25]. Interestingly, we are then able to prove that if any bipartite state is shown to be steerable via such measurements then it is also Bell nonlocal. Future work in this direction would find necessary and sufficient inequalities for more than two parties and several POVM's on each site to further illuminate the differences between steerable and Bell nonlocal states. Can the distinction be shown with two \emph{d}-outcome measurements, or do we need three measurements? What minimum measurements are required to demonstrate this distinction for higher-dimensional bipartite systems?

\textit{Note added in proof}. Recently \cite{costa2015quantification} found a related result in the restricted context where Bob's measurements used to test steering are mutually unbiased. This, however, leaves open the possibility that arbitrary qubit measurements by Bob can discriminate steering from Bell nonlocality. We establish the equivalence between steering and Bell nonlocality for the most general CHSH scenario, that is, two dichotomic measurements by both parties. More recently, the authors of \cite{quan2016steering} also showed this steering-Bell nonlocality equivalence but only for T states, i.e.~states in which $r=s=0$ in equation Eq. \eqref{eq:bipartite}, so that Alice and Bob's reduced states are mixed states. They also gave a geometric meaning of the maximum inequality violation in terms of the steering ellipsoid. Similarly to our paper they extended it to arbitrary states in \cite{quan2015einstein}, but without deriving a generalised steering inequality. 

\section*{ACKNOWLEDGMENTS} The authors acknowledge Curtis Broadbent for prompting the question on connecting Bell nonlocality and steering, and useful discussions and feedback from Cyril Branciard, Andrew Doherty, Michael Hall and Howard Wiseman. We also thank an anonymous referee for suggesting further research directions based on this work. P.G. acknowledges support from the ARC via the Centre of Excellence in Engineered Quantum Systems (EQuS), Project No. CE110001013.

\section*{APPENDIX A}
  \renewcommand{\theequation}{A\arabic{equation}}
\setcounter{equation}{0}  

We prove here a theorem that results in the necessary and sufficient EPR-steering inequality for the CHSH scenario, that is Eq. \eqref{eq:main}.

Theorem:
Let $C_1$, $C_2$ be convex sets in four dimensions and in separate planes spanned by the axes i.e. $C_1\subseteq span(\mathbf{e_1}, \mathbf{e_2})$ and $C_2 \subseteq span(\mathbf{e_3},\mathbf{e_4})$ where $\mathbf{e_1}, \mathbf{e_2}, \mathbf{e_3}, \mathbf{e_4}$ are basis vectors of four-dimensional space. If the boundaries of the sets are conic sections represented by Cartesian equations that only contain quadratic terms i.e. $C_1$ can be described as $f(x_1,x_2)=ax_1^2+bx_2^2+cx_1x_2\leq r_1$ and $C_2$ as $g(x_3,x_4)=a'x_3^2+b'x_4^2+c'x_3x_4\leq r_2$, then the convex hull \emph{C} of $C_1$ and $C_2$ has the form 
\bq 
\sqrt{f(v_1,v_2)}+\sqrt{f(v_3,v_4)}\leq max[\sqrt{r_1}, \sqrt{r_2}]
\eq

Proof:

Let $\mathbf{h_1}=(v_1,v_2,0,0)$, $\mathbf{h_2}=(0,0,v_3,v_4)$ and $\mathbf{v}=(v_1,v_2,v_3,v_4)=\mathbf{h_1}+\mathbf{h_2}$. If $\mathbf{v}$ is in the convex hull of $C_1$ and $C_2$ then $\mathbf{v}=p_1\mathbf{w_1}+p_2\mathbf{w_2}$ where $p_1+p_2=1$ and $\mathbf{w_1}$ lies in $C_1$ and $\mathbf{w_2}$ lies in $C_2$. Hence, with the assumption that $C_1\subseteq span(\mathbf{e_1}, \mathbf{e_2})$ and $C_2 \subseteq span(\mathbf{e_3},\mathbf{e_4})$ where $\mathbf{e_1}=(1,0,0,0), \mathbf{e_2}=(0,1,0,0), \mathbf{e_3}=(0,0,1,0), \mathbf{e_4}=(0,0,0,1)$:

\bq
\mathbf{w_1}=\frac{\mathbf{h_1}}{p_1}
\eq
\bq
\mathbf{w_2}=\frac{\mathbf{h_2}}{p_2}
\eq

And as $\mathbf{w_1}$ lies in $C_1$ and $\mathbf{w_2}$ lies in $C_2$:

\bq
f(\frac{v_1}{p_1},\frac{v_2}{p_1})\leq r_1
\eq
\bq
g(\frac{v_3}{p_2},\frac{v_4}{p_2})\leq r_2
\eq
Since  $f(v_1, v_2)$ and $g(v_3, v_4)$ only contain quadratic terms this implies:
\bq
\frac{1}{p_1^2}f(v_1,v_2)\leq r_1
\eq
\bq
\frac{1}{p_2^2}g(v_3,v_4)\leq r_2
\eq
Putting these together we get:
\ba
\sqrt{f(v_1,v_2)}+\sqrt{f(v_3,v_4)}&\leq p_1\sqrt{r_1}+p_2\sqrt{r_2}\\
&\leq max[\sqrt{r_1}, \sqrt{r_2}]
\end{align}
where $p_1+p_2=1$ has been used in the last line.
Applying this to our situation where the boundaries for both $C_1$ and $C_2$ have the form of Eq. \eqref{eq:quadraticproj} we obtain equation Eq. \eqref{eq:main_vbasis} as desired.

In Appendix B of \cite{cavalcanti2015analog} a proof is provided that only points in $C$ satisfy the LHV-LHS inequality in that paper. The proof can be applied in whole to inequality \eqref{eq:main_vbasis} since the inequality satisfies the properties crucial to the proof: it is of the form $f(\mathbf{v})\leq1$ where $f(\mathbf{v})$ is a convex function and its upper bound of $1$ is obtained for points  $\mathbf{v}\in C$ that can be expressed as a convex combination of a point on the boundary $\partial C_{1}$
of $C_{1}$ with a point on the boundary $\partial C_{2}$ of $C_{2}$. The latter statement is seen from the derivation above since the inequality \eqref{eq:main_vbasis} achieves the bound $1$ if and only if $f(\frac{v_1}{p_1},\frac{v_2}{p_1})=r_1$ and $g(\frac{v_3}{p_2},\frac{v_4}{p_2})=r_2$, i.e., $\mathbf{w_1}, \mathbf{w_2}$ lie on the boundaries of $C_1$ and $C_2$ respectively.

Hence only points in $C$ satisfy Eq. \eqref{eq:main_vbasis}, which in the measurement basis is Eq. \eqref{eq:main}. Thus Eq. \eqref{eq:main} is indeed the necessary and sufficient EPR-steering inequality for arbitrary measurements in the CHSH scenario. 

\section*{APPENDIX B}
  \renewcommand{\theequation}{B\arabic{equation}}
\setcounter{equation}{0}  
We examine here the case where \emph{B} and \emph{B'} are dichotomic POVM's measured by Bob. 

An arbitrary dichotomic POVM element associated with outcome 1 for observable \emph{B} can be expressed as:
\ba
E_{1|B}&=\lambda_{1|B}|1\ket\bra1|+\lambda_{2|B}|2\ket\bra 2|\\
&=k_B|1\ket\bra 1|+\lambda_{2|B}I
\end{align}
where $0\leq\lambda_{2|B}\leq\lambda_{1|B}\leq1$ are eigenvalues of $E_{1|B}$, $|1\ket$ and $|2\ket$ are corresponding orthonormal eigenstates and $k_{B}=(\lambda_{1|B}-\lambda_{2|B})$. 

Likewise for observable \emph{B'}:
\bq
E_{1|B'}=k_B'|1'\ket\bra1'|+\lambda_{2|B'}I
\eq

Decompose $|1'\ket $ in the eigenbasis of $E_{1|B}$:
\bq
|1'\ket=\sqrt{\mu}|1\ket+\sqrt{1-\mu}e^{i\phi}|2\ket
\eq
as well as the pure state received by Bob:
\bq
|\psi\ket=\sqrt{\mu'}|1\ket+\sqrt{1-\mu'}e^{i\phi'}|2\ket
\eq
Then the probability $p(1|B)$ of outcome 1 when measuring \emph{B} is:
\ba
p_{1}^{B}&=\bra \psi |E_{1|B} | \psi \ket\notag\\
&=k_{B}\mu'+\lambda_{2|B}
\end{align}
so that:
\bq
\mu'=\frac{p(1|B)-\lambda_{2|B}}{k_{B}}\label{eq:muprime}
\eq
Then for \emph{B'} the probability of outcome 1 is:
\ba
p_{1}^{B'}&=\bra \psi |E_{1|B'} | \psi \ket\notag\\
&=k_B'|\bra \psi|1'\ket|^2+\lambda_{2|B'}\notag\\\
&=k_B'[\mu'\mu+(1-\mu')(1-\mu)\notag\\
&+2\sqrt{\mu'(1-\mu')\mu(1-\mu)}\cos(\phi'-\phi)]+\lambda_{2|B'}\notag\\
&=\lambda_{2|B'}+\frac{k_B'}{k_{B}}[(p_{1}^{B}-\lambda_{2|B})\mu\notag\\
&+(k_{B}-(p_{1}^{B}-\lambda_{2|B}))(1-\mu)\notag\\
&\hspace*{-1cm}+2\sqrt{(p_{1}^{B}-\lambda_{2|B})(k_{B}-(p_{1}^{B}-\lambda_{2|B}))\mu(1-\mu)}\cos(\phi'-\phi)]\label{eq:ellipse2}
\end{align}

Now let $y=p_{1}^{B'}, x=p_{1}^{B}, \alpha=\lambda_{2|B'}, \beta=\frac{k_B'}{k_B}, \gamma=\lambda_{2|B}, \delta=k_B+\lambda_{2|B}$. The boundary of the curve $p_{1}^{B'}$ versus $p_{1}^{B}$ according to Eq. \eqref{eq:ellipse2} is achieved with $\cos(\phi'-\phi)=\pm1$ i.e. $\cos(\phi'-\phi)^2=1$.
Then the boundary has the form:
\ba
y&=\alpha+\beta[(x-\gamma)\mu+(\delta-x)(1-\mu)\notag\\
&\pm2\sqrt{(x-\gamma)(\delta-x)\mu(1-\mu)}\cos(\phi'-\phi)\notag\\
&=\alpha+r(x-\gamma)+s(\delta-x)\pm t\sqrt{(x-\gamma)(\delta-x)}
\end{align}
where $r=\beta\mu, s=\beta(1-\mu), t=2\beta\sqrt{\mu(1-\mu)}$.

Then rearranging:
\ba\label{eq:curve1}
((s-r)^2&+t^2)x^2 +2(s-r)xy+y^2\notag\\
&+(2(s-r)(r\gamma-s\delta-\alpha)-t^2(\delta+\gamma))x\notag\\
&+2(r\gamma-s\delta-\alpha)y+((r\gamma-s\delta-\alpha)^2+t^2\gamma\delta)=0
\end{align}
let:
\ba
A&=(s-r)^2+t^2\notag\\
&=\beta^2(1-2\mu)^2+4\beta^2\mu(1-\mu)\notag\\
&=\beta^2\\
B&=(s-r)\notag\\
&=\beta(1-2\mu)\\
C&=1\\
D&=(s-r)(r\gamma-s\delta-\alpha)-\frac{t^2(\delta+\gamma)}{2}\\
F&=r\gamma-s\delta-\alpha\\
G&=(r\gamma-s\delta-\alpha)^2+t^2\gamma\delta
\end{align}
So Eq. \eqref{eq:curve1} is in the form:
\bq\label{eq:quadratic}
Ax^2+2Bxy+Cy^2+2Dx+2Fy+G=0
\eq

Equation \eqref{eq:quadratic} describes an ellipse in terms of the variables \emph{x, y} i.e., $p_{1}^{B}, p_1^{B'}$. We will now calculate its semi-axis lengths, centre and counterclockwise angle of rotation from the \emph{x} axis to the major axis based on the formulas in \cite{weissteinellipse}.

$X_{C}$, the \emph{x} coordinate of the ellipse centre, is:
\ba
X_{C}&=\frac{CD-BF}{B^2-AC}\notag\\
&=\frac{\frac{-t^2(\delta+\gamma)}{2}}{(s-r)^2-((s-r)^2+t^2)}\notag\\
&=\frac{1}{2}(\delta+\gamma)\notag\\
&=\frac{k_B}{2}+\lambda_{2|B}
\end{align}
For projective measurements $k_B=1, \lambda_{2|B}=0$ which implies $X_{C}=\frac{1}{2}$ as expected.

$Y_{C}$, the \emph{x} coordinate of the ellipse centre, is:
\ba
Y_{C}&=\frac{AF-BD}{B^2-AC}\notag\\
&=\frac{t^2(r\gamma-s\delta-\alpha)+\frac{(s-r)(\delta+\gamma)t^2}{2}}{(s-r)^2-((s-r)^2+t^2)}\notag\\
&=-[\frac{(r+s)(\gamma-\delta)}{2}-\alpha]\notag\\
&=\frac{k_B'}{2}+\lambda_{2|B'}
\end{align}
For projective measurements $k_B'=1, \lambda_{2|B'}=0$ which implies $Y_{C}=\frac{1}{2}$ as expected.

The semi-axis lengths are given by:
\ba\label{eq:axis}
a_{\pm}=\sqrt{\frac{2(AF^2+CD^2+GB^2-2BDF-ACG)}{(B^2-AC)[\pm\sqrt{(A-C)^2+4B^2}-(A+C)]}}
\end{align}
Now we can write:
\ba
A&=B^2+t^2\\
D&=BF-\frac{t^2l}{2}\\
G&=F^2+t^2l'
\end{align}
where $l=\delta+\gamma, l'=\delta\gamma$.
The numerator under the square root of Eq. \eqref{eq:axis} is then:
\ba
2[(B^2+t^2)F^2&+(BF-\frac{t^2l}{2})^2+(F^2+t^2l')B^2\notag\\
&-2BF(BF-\frac{t^2l}{2})-(B^2+t^2)(F^2+t^2l')]\notag\\
&=2t^4(\frac{l^2}{4}-l')\notag\\
&=2t^4(\frac{(\delta+\gamma)^2}{4}-\delta\gamma)\notag\\
&=2t^4k_B^2
\end{align}
And the denominator under the square root of Eq. \eqref{eq:axis} is:
\ba
&(B^2-(B^2+t^2))[\pm\sqrt{(B^2+t^2-1)^2+4B^2}-(B^2+t^2+1)]\notag\\
&=-t^2[\pm\sqrt{(\beta^2+1)^2+16\beta^2\mu(\mu-1)}-(\beta^2+1)]\notag\\
&=-t^2[\pm\sqrt{((\frac{k_B'}{k_B})^2+1)^2+16(\frac{k_B'}{k_B})^2\mu(\mu-1)}\notag\\
&-((\frac{k_B'}{k_B})^2+1)]
\end{align}
Hence, substituting the value for t:
\bq
a_{\pm}=2\frac{k_B^2}{k_B'}\sqrt{\mu(1-\mu)}\sqrt{\frac{-2}{S}}
\eq
where:
\bq
S=\pm\sqrt{((\frac{k_B'}{k_B})^2+1)^2+16(\frac{k_B'}{k_B})^2\mu(\mu-1)}-((\frac{k_B'}{k_B})^2+1)
\eq
For projective measurements $k_B=k_B'=1$ so:
\ba
a_{\pm}&=2\sqrt{\mu(1-\mu)}\sqrt{\frac{-2}{\pm2\sqrt{1+4\mu(\mu-1)}-2}}\notag\\
&=2\sqrt{\mu(1-\mu)}\sqrt{\frac{-1}{\pm\abs{2\mu-1}-1}}
\end{align}
If $\mu=0.5$ then $a_{\pm}=\sqrt{\frac{-1}{-1}}=1$ as expected.

The counterclockwise angle of rotation from the \emph{x} axis to the major axis is:
\ba
\phi&=\frac{1}{2}\cot^{-1}(\frac{A-C}{2B})\notag\\
&=\frac{1}{2}\cot^{-1}(\frac{\beta^2-1}{2\beta(1-2\mu)})\notag\\
&=\frac{1}{2}\cot^{-1}(\frac{k_B'^2-k_B^2}{2k_Bk_B'(1-2\mu)})
\end{align}
For projective measurements and $\mu\neq0.5$ we get  $\phi=\frac{1}{2}\cot^{-1}(0)=\frac{\pi}{4}$.

The general parametric form of the ellipse in terms of the above is:
\ba
x&=X_{C}+a\cos(\xi)\cos(\phi)-b\sin(\xi)\sin(\phi)\notag\\
&=X_{C}+T\cos(\xi+\kappa)\\
y&=Y_{C}+a\cos(\xi)\sin(\phi)+b\sin(\xi)\cos(\phi)\notag\\
&=Y_{C}+T'\cos(\xi+\kappa')
\end{align}
where
\ba
T&=\sqrt{a^2\cos^2(\phi)+b^2\sin^2(\phi)}\\
\kappa&=\tan^{-1}(\frac{b}{a}\tan(\phi))\\
T'&=\sqrt{a^2\sin^2(\phi)+b^2\cos^2(\phi)}\\
\kappa'&=\tan^{-1}(\frac{a}{b}\tan(\phi))
\end{align}
This implies:
\ba
2p_{1}^{B}(\xi)-1 & = 2(X_{C}+T\cos(\xi+\kappa)-\frac{1}{2})\\
2p_{1}^{B'}(\xi)-1 & =2(Y_{C}+T'\cos(\xi+\kappa')-\frac{1}{2})
\end{align}

Then the vectors making up the boundaries of $C_{1}$ to $C_{4}$ in the basis \eqref{eq:basis} have the form:
\ba
C_1&: 2(X_{C}+T\cos(\xi+\kappa)-\frac{1}{2})\mathbf{e}_{1}\notag\\
&+2(Y_{C}+T'\cos(\xi+\kappa')-\frac{1}{2})\mathbf{e}_{2}\\
C_2&: 2(X_{C}+T\cos(\xi+\kappa)-\frac{1}{2})\mathbf{e}_{3}\notag\\
&+2(Y_{C}+T'\cos(\xi+\kappa')-\frac{1}{2})\mathbf{e}_{4}\\
C_3&: -(2(X_{C}+T\cos(\xi+\kappa)-\frac{1}{2})\mathbf{e}_{1}\notag\\
&+2(Y_{C}+T'\cos(\xi+\kappa')-\frac{1}{2})\mathbf{e}_{2})\\
C_4&:-(2(X_{C}+T\cos(\xi+\kappa)-\frac{1}{2})\mathbf{e}_{3}\notag\\
&+2(Y_{C}+T'\cos(\xi+\kappa')-\frac{1}{2})\mathbf{e}_{4})
\end{align}

Each of these correlation boundaries are elliptical and for the projective case ($A=X_{C}=\frac{1}{2}$) $C_1$ and $C_3$ reduce to $\cos(\xi+\kappa)\mathbf{e}_{1}+ \cos(\xi-\kappa)\mathbf{e}_{2}$ and $C_2$ and $C_4$ to $\cos(\xi+\kappa)\mathbf{e}_{3}+ \cos(\xi-\kappa)\mathbf{e}_{4}$ as we have seen before. The equation of the boundary of $C_1$ and $C_3$ can be found as follows:
Let $m=2p_{1}^{B}(\xi)-1 =2x-1$ i.e., $x=\frac{m+1}{2}$ and $n=2p_{1}^{B'}(\xi)-1 =2y-1$, i.e., $y=\frac{n+1}{2}$. Then from Eq. \eqref{eq:quadratic} we get
\ba
A(\frac{m+1}{2})^2&+2B(\frac{m+1}{2})(\frac{n+1}{2})+C(\frac{n+1}{2})^2\notag\\
&+2D(\frac{m+1}{2})+2F(\frac{n+1}{2})+G=0
\end{align}
So,
\ba\label{eq:quadraticmn}
(\frac{A}{4})m^2&+(\frac{C}{4})n^2+(\frac{B}{2})mn+(\frac{A+B}{2}+D)m\notag\\
&+(\frac{B+C}{2}+F)n=-(\frac{A+C}{4}+\frac{B}{2}+D+F+G)
\end{align}
For projective measurements this reduces to Eq. \eqref{eq:quadraticproj}.
The equation for $C_2$ and $C_4$ involves replacing $m$ by $-m$ and $n$ by $-n$.

For the general POVM case $C_1$ and $C_3$ lie in the same plane but are distinct sets as with $C_2$ and $C_4$. The convex hull \emph{C} of the sets is the convex hull of $C_5$ and $C_6$ where $C_5$ is the convex hull of $C_1$ and $C_3$ and $C_6$ is the convex hull of $C_2$ and $C_4$. The boundary of $C_5$ consists of the two outer common tangents to the ellipses $C_1$ and $C_3$ and the outer arcs of the ellipses that connect with the tangents, and $C_6$ has the same equation for its boundary as $C_5$ (but in an orthogonal plane). The boundary is piecewise defined so that \emph{C} cannot be expressed as a simple inequality for the general POVM scenario.
\vspace{8cm}
\bibliography{NecFINAL_arxivV2}

\end{document}